# Molecular Plasmonics: strong coupling at the low molecular density limit

Lihi Efremushkin, Maxim Sukharev and Adi Salomon*

Arizona State University, Mesa, AZ 85212, USA

*Department of Chemistry, Institute of Nanotechnology and Advanced Materials (BINA), Bar-Ilan University, Ramat-Gan 5290002, Israel.

E-mail: Adi.Salomon@biu.ac.il

## Abstract

We study the strong coupling between the molecular excited state and the plasmonic modes of silver hole arrays with a resonant frequency very close to the asymptotic line of the plasmonic dispersion relation, at the nonlinear regime. We demonstrate that the strong coupling regime can be achieved between the two sub-systems at low molecular densities with negligible damping of the electromagnetic field. Our results are supported by rigorous numerical simulations showing that the strong coupling is observed when the molecular transition lies within the nonlinear regime of the dispersion relation rather than the linear regime.

## Introduction

Strong interaction between photonic modes and molecular states can lead to hybrid materials with new photophysical properties.[1–5] In the strong coupling regime, damping rates are small compared to the light-matter interaction, and the exchange of energy between excited molecular states and photon states occurs on a very short time scale. Such interactions are expected to modify the dynamics of the molecular system and may lead to hybrid materials with unique optoelectronic properties.[4–9] When optically coupled molecules and photonic modes form hybrid states, new pathways for energy redistribution might be opened and thus can lead to new photochemistry occurring on metallic surfaces. These perspectives provide a new way to tailor the physical properties of a given molecular system, namely, to modify the molecular surface potential energy and control the energy-redistribution pathways, electron-transfer processes and radiative emission. For example, if coupling to the radiation field occurs, a change in the molecular emission can be observed; thus, when a coherent mixed exciton-plasmon (or exciton-cavity mode) is formed, a collective response of the whole system occurs. This is evident from the dependence of the Rabi splitting values associated with the exciton-plasmon coupling on the molecular density.[10] Traditionally, strong coupling has been studied using micro-cavity resonators and other solid-state systems,[2] which demand cryogenic temperatures and state-of-the-art fabrication. Recently, the strong coupling regime has been realized with metallic nanostructures in which the plasmonic modes, coherent oscillation of the metal's free electrons, replace the photonic modes of the microcavities.[1,7,11,12] These metallic nanostructures have been found to reach the strong coupling regime at room temperature due to the confinement of the electromagnetic field to a deep subwavelength volume. The strong confinement of these photonic (plasmonic) modes and the ability to control their propagation and localization on the surface at any given optical frequency are the main reasons to study their interaction with molecules.

So far, the strong coupling regime has been observed using subwavelength arrays and metallic nanoparticles.[1,10,11,13–17] Most of the experimental reports are mainly on an ensemble of excitons (molecules or quantum dots) and nanoparticles. Recently, the strong coupling regime has been realized using bowtie nanoantennas and semiconductor quantum dots.[18] The enhanced electromagnetic field trapped in the very small volume between the particles is responsible for achieving the single quantum-emitter limit. The strong coupling regime was also realized using single molecules, where similar linear optical measurements were observed.[19] Still, there are ambiguities in the interpretation of the optical linear measurements, which are the predominant method to observe the strong coupling regime in plasmonic systems. Specifically, the observed splitting at the molecular level may be due to enhanced absorption rather than strong interaction between excitons and plasmons.[10]

In most studies, the strong coupling regime is observed using a relatively high concentration of J-aggrerate molecules,[1,9,11,13,16,20–23] and the color of the thin layer can be seen by a naked eye. In these cases, changes in the refractive index due to molecular dipole transitions also play a role in the observed spectrum. The molecules are phyically adsorbed onto the metallic structure (spin-coated or drop-casted). Clearly, both the oscillator strength and the high concentration are critical to reaching the strong coupling regime at room temperature.

We note that strong Rabi splitting values up to about 700 meV have been observed,[24] which is about 30% of the molecular transition-state energy, yet, no clear change in the nature of the molecular system has been observed. Only a small number of studies show a change in the nature or dynamic behavior of the molecular system due to strong coupling to plasmonic modes.[1,5,7,9] It is reasonable to assume that in such systems, only some of the molecules strongly interact with the plasmomic modes, whereas the other molecules are weakly coupled and their interaction results in the dumping of the transmitted light.

Here, we reach the strong coupling regime using a relatively low molecular concentration of 0.2 mM and an absorbance of about 0.015, an order of magnitude lower than those used in other studies. The reason we could achieve the strong coupling at such a low molecular concentration is that our molecular system has a transition state that lies whithin the nonlinear region of the plasmonic dispersion, very close to the asymptotic line (see figure 1). In that region, light–matter interaction is significantly enhanced and small modification in the refractive index of the dielectric medium that is in contact with the plasmonic system should lead to strong modulations in the dispersion.[25]

Figure 1 shows a modulation of the surface plasmons (SPs) dispersion relation following the deposition of the molecular layer with a relatively sharp transition state at various frequencies. We examine three types of molecules with the same physical properties, such as oscillator strength, concentration and lifetime, but with different transition states (figure 1a). A very strong modulation of the dispersion relation is observed when the molecular transition lies near 420 nm, a moderate one when the molecular transition is at 530 nm and a smaller one when the transition lies at 620 nm. In figure 1b, the group velocity, $V_{group} = d\omega / dk$, is very small and, therefore, in this case, a strong light–matter interaction is expected.

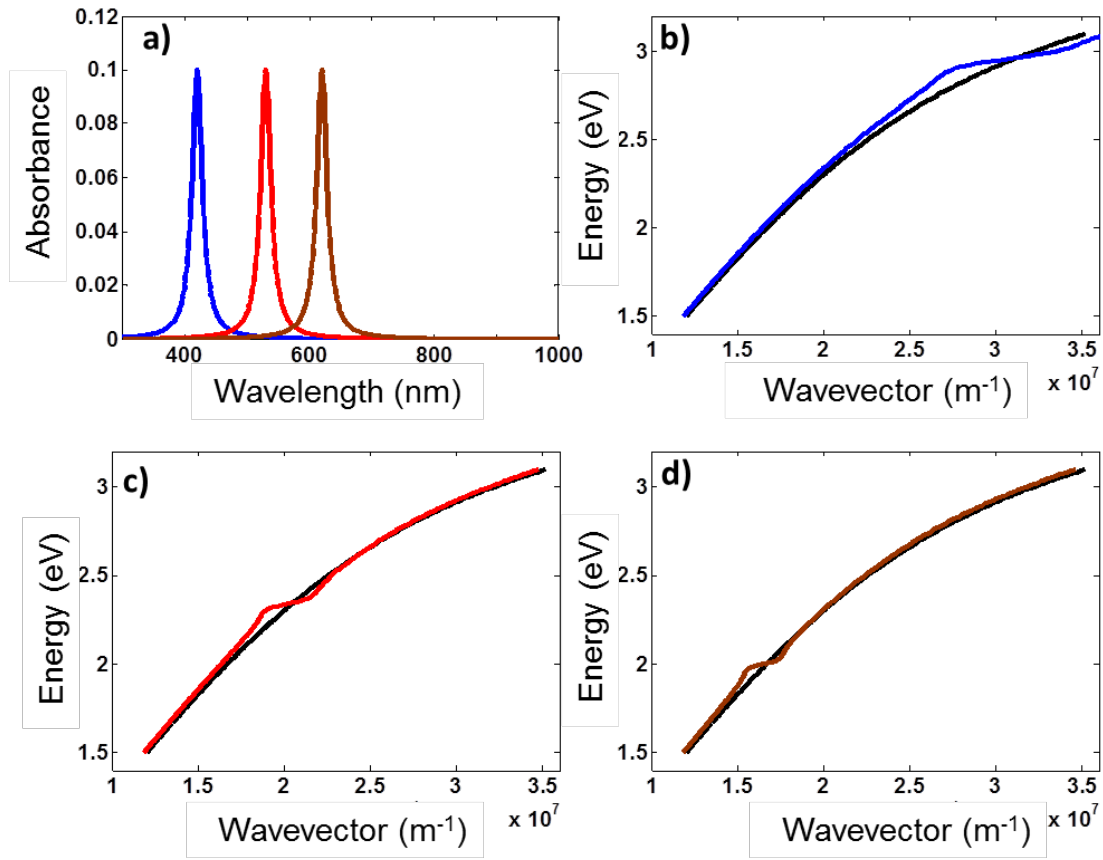

**Figure 1:** (a) simulated absorbance spectrum of dye molecules $H_2TPPS_4$ with an excited state at 420 nm (blue line), at 530 nm (red line) and at 620 nm (brown line). (b-d) Dispersion curves of surface plasmons of a flat semi-infinite silver in one dimension: black line is for metal covered by a dispersionless polymer, blue line corresponds to metal covered by dye molecules resonant at 420 nm (blue line in (a)), red line is for molecules resonant at 530 nm (red line in (a)), and brown line is for molecules resonant at 620 nm (brown line in (a)).

Following the explanation above, we chose to work with tetraphenylporphyrin tetrasulfonic acid ($H_2TPPS_4$), a phorphyrin derivative monomer whose absorption band is at ~3 eV (420 nm) with an absorption coefficient of $\varepsilon = 7\times10^6$ $M^{-1}cm^{-1}$. The absorption spectrum of a thin layer of 0.23 mM (for correlation to density see table S1) $H_2TPPS_4$ embedded in polyvinyl alcohol (PVA) is shown in figure 2a. We deposited those molecular layers onto hexagonal silver hole arrays and studied their interaction with the molecular transition state. Figure 2b shows a schematic illustration of the device we used for this study. A layer of H$_2$TPPS$_4$-doped PVA is spin-coated on a hexagonal silver hole array.

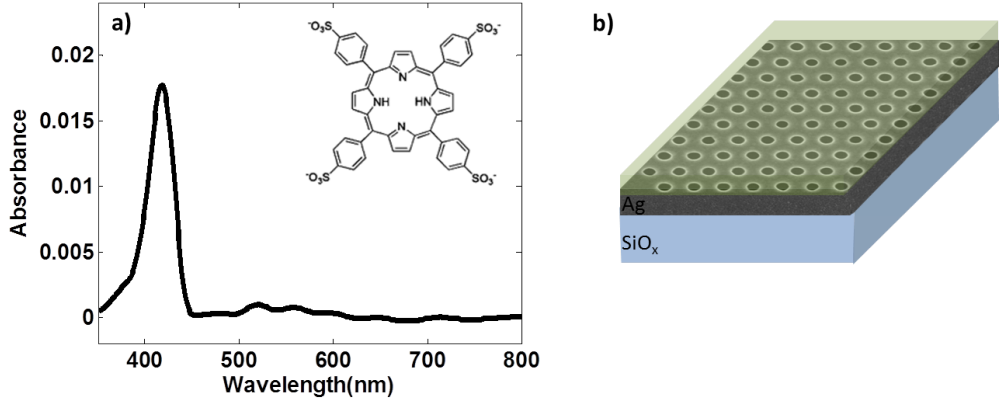

**Figure 2: (a)** Absorbance spectrum of H$_2$TPPS$_4$-doped PVA spin-coated onto glass. The peak position of the S$_5$ exciton band is at ~3 eV (420 nm). The absorbance is ca. 0.016. Inset: H$_2$TPPS$_4$ molecule structure. **(b)** Schematic illustration of the system used. The system is composed of fabricated Ag hole array placed between SiO$_x$ and a layer of PVA or H$_2$TPPS$_4$-doped PVA (green layer).

Following previous studies[11,26], we chose to work on plasmonic modes of hexagonal hole arrays. The coupling to the free-propagating light is represented by Bragg scattering, as shown by the equation[27–29]

$$\left|\vec{k}_{sp}\right| = \left|\vec{k}_\parallel + i\vec{G}_x + j\vec{G}_y\right|, \quad (1)$$

where $i$ and $j$ are integers, $k_\parallel$ is the incident light wave vector, $k_{sp}$ is the SPs wave vector and $G_x$ and $G_y$ are the reciprocal lattice vectors for which $|G_x| = |G_y| = 2\pi/p$, where $p$ is the lattice period.

For this array symmetry, the peak positions at a specific wavelength $\lambda$ are described, to a first approximation, by the following expression[27–29]

$$\lambda_{max} = \frac{p}{\sqrt{\frac{4}{3}(i^2 + i\cdot j + j^2)}} \sqrt{\frac{\varepsilon_m \varepsilon_d}{\varepsilon_m + \varepsilon_d}}, \quad (2)$$

where $\varepsilon_m$ and $\varepsilon_d$ are the permittivity of the metal and the dielectric material in contact, respectively, and the integers $i$ and $j$ are the scattering orders. According to equation (2), tuning plasmonic modes to certain frequencies can easily be attained by changing the array periodicity $p$ and, therefore, can provide a systematic tool to study their interaction with molecular transition states.

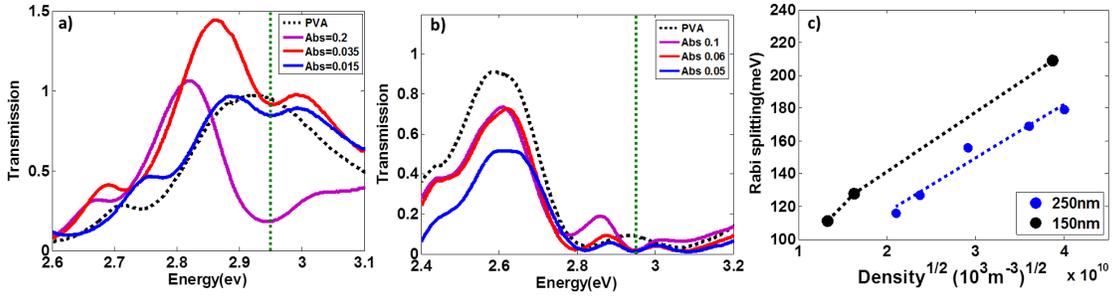

**Figure 3:** Transmission spectra of the hybrid system which consists of a hexagonal hole array covered with $H_2TPPS_4$-doped PVA with an Ag film thickness of **(a)** 150 nm and **(b)** 250 nm. The green line is a guiding line to indicate the molecular transition state **(c)** Rabi-splitting values as a function of the square root of the $H_2TPPS_4$ density for two types of Ag film thickness, blue dot are for 250 nm thick film and black dots are for 150 nm thick film.

Figures 3a and 3b show the transmission spectrum of the hexagonal hole array in which the plasmonic mode at normal incidence is at about ~3 eV. This mode is due to the coupling of the SPs modes propagating in the second Brillouin zone. When depositing molecules, the transmission spectrum of the plasmonic device changes dramatically. The original peak (black line) splits into two new eigenmodes, the lower and upper polariton, and a broadening of the peak is observed. Interestingly, almost no damping of the electromagnetic field is observed in this region, and a small to moderate enhancement of the plasmonic mode is even indicated. Notably, even if damping does occur at the lower energy regime (figure 3b at ~2.6 eV), a splitting to two new eigenmodes is observed. The splitting is nonsymmetric, whereas the lower polariton gives rise to a moderate enhancement in the transmission.

We repeated the experiment with several molecular concentrations and hole arrays of different thickness and periodicities, and we found that the Rabi splitting values scale linearly with the square root of the molecular density, as expected (figure 3c).[2-3] The strong coupling in our studies is achieved at very low molecular concentrations compared to other studies with similar plasmonic system and for nanoparticles (Table 1). For example, Sugawara *et al.*[13] observed a splitting of about 230 meV in a very concentrated molecular film with relatively high optical density (absorbance of 0.9). This is not surprising, since the molecular transition state lies at 1.85 eV, in the linear regime of the dispersion relation. Indeed, when approaching the nonlinear regime, with the molecular transition state lying at ~2.5 eV as in the case of Salomon *et al.*[11], the optical density of the molecular film is lower (absorbance of 0.46) and the same Rabi-splitting values of 217 meV are observed.

**Table 1:** Comparison of Rabi splitting values of different molecular plasmonic systems. For comparison with our system, the estimated Rabi splitting for absorbance of 0.2 are in brackets

| | Plasmonic system | Molecular transition $\Omega$ (eV) | Absorbance | Rabi splitting value (meV) | Ref |
|---|---|---|---|---|---|
| Silver hole array | **Hexagonal** | **2.9** | **0.2** | **210** | Current study |
| | Hexagonal | 2.53 | 0.46 | 312 (205) | [11] |
| | Square | 2.53 | 0.46 | 217 (143) | [11] |
| | Square | 1.78 | 0.8 | 250 (125) | [21] |
| Nanoparticles | Gold nanorods | 1.99 | 0.2 | 190 | [22] |
| | Silver nanoprisms | 2.11 | 2 | 295 (93) | [23] |
| | Spherical nanovoid arrays of nanostructured gold | 1.85 | 0.9 | 230 (108) | [13] |

The experimental results were simulated using homebuilt finite-difference time-domain codes that numerically integrate coupled Maxwell-Bloch equations. The details of the simulations are discussed thoroughly in the Theory section. The model is based on rate equations and can only be used to describe $H_2TPPS_4$ qualitatively. Therefore, our major goal on the modeling side was to capture the essential physics shown in the experiments. First, we simulated the transmission spectra of a set of hexagonal hole arrays covered only with a PVA layer. The transmission and the reflection spectra are shown in figures 4a and 4b. The transmission peak at 3.032 eV for the period of 330 nm corresponds to the experimental results for the 360 nm period (see Fig. S1). The corresponding steady-state spatial distributions of the electromagnetic intensity are shown in figures 4c and 4d. The energy of the mode with the frequency near 3 eV is highly localized inside the PVA layer near the metal surface, making coupling to the molecules highly efficient. The modes observed at 2.65 eV and 3.55 eV in the transmission spectrum for the period of 330 nm are more spatially delocalized and, therefore, result in a noticeably smaller Rabi splitting. All three modes correspond to surface plasmon-polariton resonances of a different spatial character, since they all appear as maxima in the transmission and minima in the reflection.

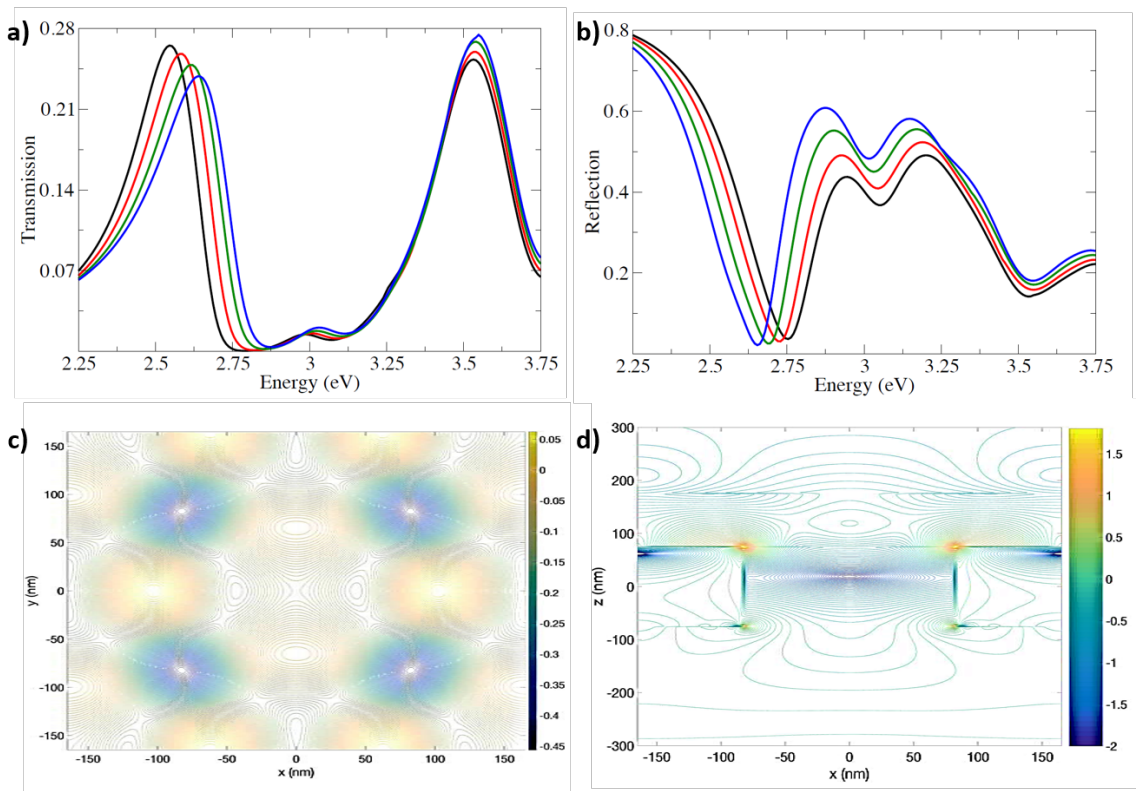

**Figure 4:** Transmission (a) and reflection (b) spectra calculated for the periodic array of holes for the periods of 320 nm (black line), 330 nm (red line), 340 nm (green line) and 350 nm (blue line). The hole diameter is 166 nm, the thickness of the silver is 150 nm and the PVA layer on the input side is 100 nm thick. The output side (substrate) is simulated as semi-infinite, nondispersive dielectric with a refractive index of 1.5 corresponding to glass. Panels (c) and (d) show the electromagnetic intensity distributions (normalized with respect to the incident intensity) shown on a log scale for the resonance of 3.032 eV and period of 330 nm. The incident field propagates along the z-axis from top to bottom in panel (d) and is horizontally polarized. The transverse distribution shown in panel (c) is evaluated inside the PVA layer at 50 nm above the metal.

It is important to emphasize that the high localization of the fields inside the PVA layer is due to the fact that the PVA and the substrate have the same refractive index. We varied the refractive index of the PVA in order to see how this affects the mode and found that higher and lower indexes of the PVA compared to the glass result in a significantly broader mode with lower local field enhancement. Investigations that are more thorough are currently under way and are the subject of another manuscript that we are currently working on.

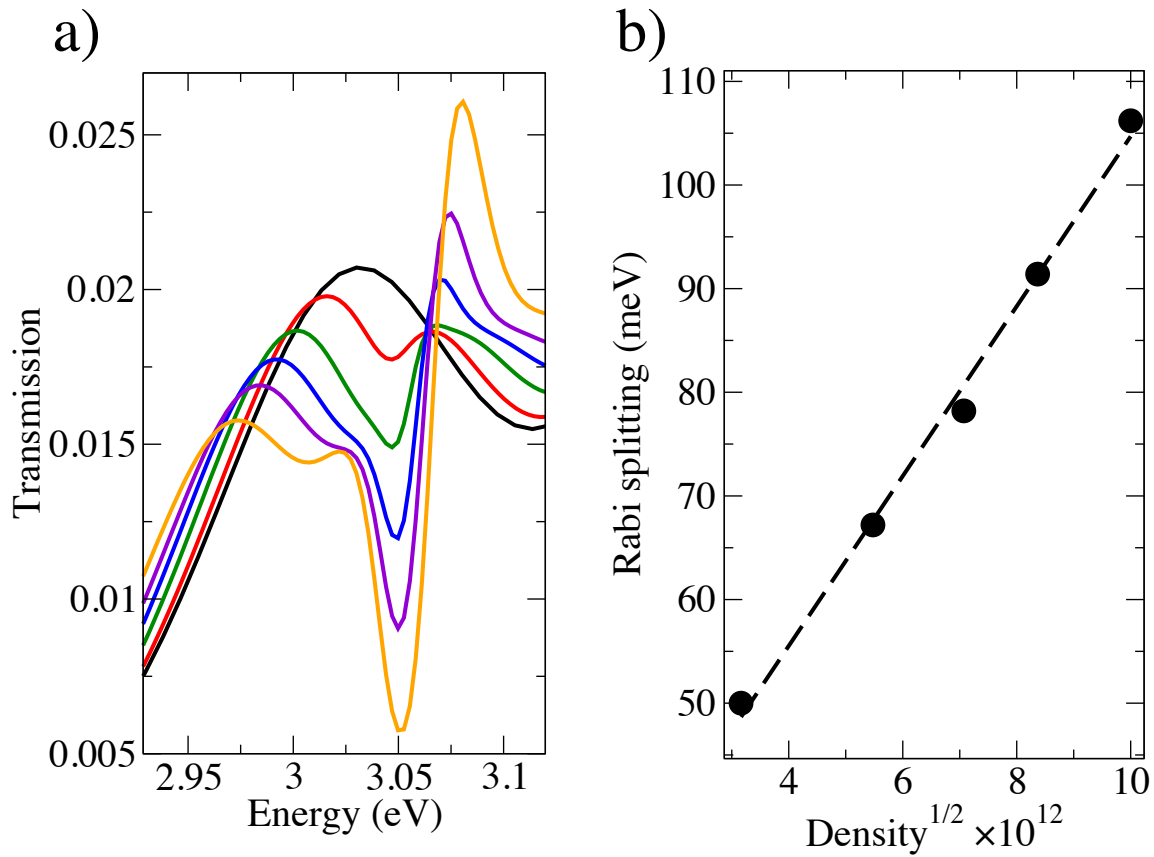

**Figure 5:** Panel **(a)** shows transmission as a function of the incident photon energy calculated for the period of 330 nm. Transmission with no molecules in PVA is shown as a black line. Simulations with molecules doped in PVA for molecular concentrations are $10^{25}$ m$^{-3}$ (red line), $3\times10^{25}$ m$^{-3}$ (green line), $5\times10^{25}$ m$^{-3}$ (blue line), $7\times10^{25}$ m$^{-3}$ (violet line) and $10^{26}$ m$^{-3}$ (orange line). Panel **(b)** shows the values of the Rabi splitting extracted from the transmission spectra as a function of the square root of the number density (m$^{-3}$)$^{1/2}$. The dashed line shows the linear fitting. Other parameters used in the simulations are the molecular transition energy (3.05 eV), the transition dipole moment (10 Debye), the radiationless lifetime of the excited state ($4.14\times10^{-3}$ eV), the pure dephasing rate ($1.25\times10^{-2}$ eV), the PVA thickness (100 nm), the silver thickness (150 nm) and the hole diameter (166 nm).

We now turn to the strong coupling of molecules to the mode near 3 eV. In our simulations (see figure 5), every induced dipole contributes to the coupling with a local electric field. We, therefore, concentrate on a qualitative comparison of the strong coupling with plasmon modes supported by hole arrays. The molecular resonance is set at a slightly higher value than the plasmon mode (as in the experiments), and the dependence of the Rabi splitting on the molecular concentration is investigated. Figure 5a shows the transmission spectra at various molecular concentrations. The strong coupling is clearly observed for all densities. For those molecular parameters specified in the figure caption, the strong coupling is achieved at densities of $1 \times 10^{25}$ m$^{-3}$ and greater.[7] Moreover, we see the appearance of a third mode in the spectra at high molecular concentrations. The physics of this mode is related to the strong dipole–dipole interactions enhanced by the plasmon field, and is discussed by us elsewhere.[7] Upon extracting the values of the energies of the upper and lower polariton modes and plotting their differences as a function of the molecular concentration (figure 5b), we observe the conventional square-root dependence.[3] Making molecules resonate with lower energy plasmon resonances (2.33 eV) results in noticeably lower values of the Rabi splitting, indicating that the plasmon mode in the nonlinear regime (3.032 eV) indeed leads to stronger coupling (see Fig S2).

Finally, we study the strong coupling when modifying the SP modes to be on/off resonance with respect to the molecular transition states. The hole-array periodicity varies between 330 and 450 nm, where the hole diameter d = 166 nm is kept constant.

Since the molecular transition state is very close to the asymptotic line, the plasmonic modes were aligned to be either resonant or lower in energy with respect to the molecular transition state. The transmission spectra of the hexagonal hole array covered by PVA are shown in figure 6a. The transmission peak becomes clearly red-shifted as the periodicity increases, as expected for a hexagonal hole array. Figure 6b shows the transmission spectra of the hybrid system, the hexagonal hole array, covered with 100 nm of $H_2TPPS_4$ doped in PVA. The molecular layer adsorption is about 0.015, which indicates a molecular density of $1.8 \times 10^{20}$. A clear splitting is observed at the molecular level. When the SP modes are resonant (blue curve), the splitting is symmetric, whereas when they are off resonant, the splitting is asymmetric. At a periodicity of 400 nm, the higher polariton is barely observed, which is consistent with the fact that it is very close to the asymptotic line.

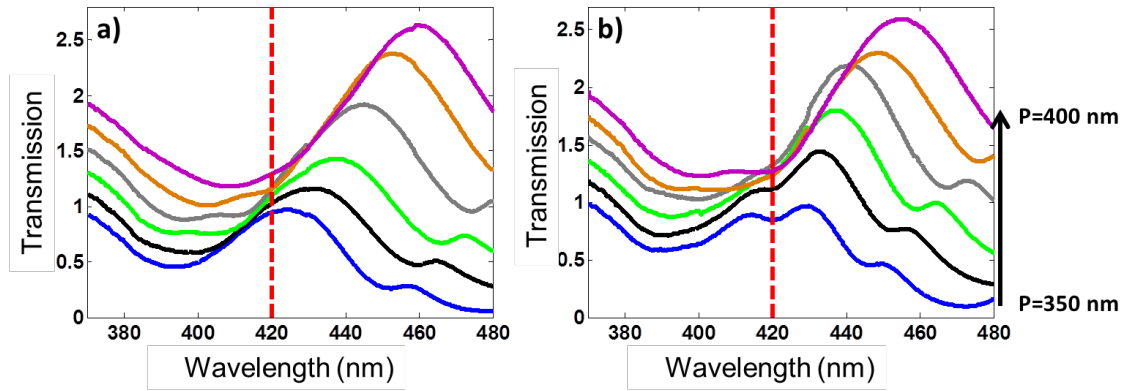

**Figure 6:** Transmission spectra of hexagonal hole array of Ag with periodicity varying between 350 (blue) and 400 nm (pink) (increments of 10 nm), coated with **(a)** PVA and **(b)** $H_2TPPS_4$-doped PVA. The thickness of the Ag film is 150 nm. The vertical line at 420 nm (~3 eV) is for the $S_5$ absorption band. By changing the periodicity, the transmission peak can be tuned to be on/off resonance with respect to the molecular exciton. When the transmission peak is resonant, splitting in the SP mode occurs when the surface is coated with the molecular system, indicating strong coupling between the molecular system and the SP modes.

In conclusion, we show that strong interaction between plasmonic modes and molecular transition states is enhanced when the molecular transition lies in the nonlinear regime of the dispersion relation. Therefore, even very low molecular density can reach the strong coupling regime. The coupling results in the formation of hybrid states that are different from the original uncoupled modes with respect to their energy levels, dispersion and dynamics. Therefore, it should be possible to use strong coupling to tune the properties of a material for a given application in areas such as photochemistry and optoelectronic devices.[30]

## Experimental section

### Sample preparation

Sub-wavelength hexagonal hole arrays were milled by a focused ion beam (FIB, Helios 600, FEI) in sputtered silver films of 150 nm and 250 nm thickness on glass substrates. The diameter of the holes was kept constant at 166 nm as the array period varied from 350 to 400 nm (increments of 10 nm). 100-nm thick PVA (89,000–98,000

Mw, Aldrich) with and without the dye molecule $H_2TPPS_4$ (4, 4′, 4″, 4‴-(Porphine-5, 10, 15, 20-tetrayl) tetrakis (benzenesulfonic acid) tetrasodium salt hydrate (Aldrich)) in basic form, pH = 6–7, was spin-coated on the sample.

Optical measurements

Spectral measurements and optical imaging of the hole arrays and molecular layer were conducted using an inverted light microscope in transmission mode (IX83, Olympus). The illumination source was a halogen lamp (100 W, Olympus). The spectroscopic data were obtained using a spectrograph (IsoPlane SCT320, Princeton Instruments), the 'LightField' program and a CCD camera (PIXIS 1024, Princeton Instruments). When using the spectrograph, the measurements were performed with the X20 objective (N.A. 0.25) and the 500-nm blaze gratings (grating density 1200 g/mm and 300 g/mm) in the spectral range of 370 to 480 nm and 380 to 520 nm respectively.

The UV-vis spectra, for the $TPPS_4$-covered glass, were collected by a CARY bio-100 spectrophotometer. PVA-coated glass was used as the reference.

## Theory section

Optical properties of a hole array schematically depicted in figure 1 are simulated using the finite-difference time-domain approach.[31] We numerically integrate Maxwell's equations

$$\mu_0 \frac{\partial \vec{B}}{\partial t} = -\nabla \times \vec{E},$$
$$\varepsilon_{\text{eff}} \varepsilon_0 \frac{\partial \vec{E}}{\partial t} = \frac{1}{\mu_0} \nabla \times \vec{B} - \vec{J}, \tag{0.0}$$

where $\vec{J}$ corresponds to the current density in the metal regions or the polarization density in the spatial regions occupied by molecules. The dielectric function of silver is taken in the form of the Drude function

$$\varepsilon(\omega) = \varepsilon_r - \frac{\Omega_p^2}{\omega^2 + i\Gamma\omega}, \tag{0.0}$$

where[32] $\varepsilon_r = 8.926$, $\Omega_p = 11.585$ eV, $\Gamma = 0.203$ eV, and $\varepsilon_{\text{eff}} = \varepsilon_r$ in (0.0). To account for the dispersion (0.0) in the time domain, we follow the conventional auxiliary differential

equation method,[33] leading to the following time-dependent equation on the density current:

$$\frac{\partial \vec{J}}{\partial t} + a\vec{J} = b\vec{E}.\tag{0.0}$$

To account for the optical response of molecules doped in PVA, we propagate a set of rate equations describing interacting two-level emitters:[6]

$$\frac{dn_2}{dt} = \frac{1}{\hbar \Omega_0}(\vec{E} \cdot \vec{P}) - \gamma_1 n_2,$$
$$\frac{d^2\vec{P}}{dt^2} + \gamma_2 \frac{d\vec{P}}{dt} + \Omega_0^2 \vec{P} = \sigma(2n_2 - n_0)\vec{E},\tag{0.0}$$

where $n_2$ is the number density of the excited molecules, $\gamma_1$ is the radiationless lifetime of the excited state, $\Omega_0$ is the molecular transition energy, $\gamma_2 = \gamma_1 + 2\gamma_{dp}$, where $\gamma_{dp}$ is the pure dephasing rate, $n_0$ is the total number density of the molecules, $\vec{P}$ is the macroscopic polarization and the coupling parameter $\sigma = \frac{2\Omega_0 d_0^2}{3\hbar}$ with the molecular transition dipole $d_0$.[34] The coupling between Maxwell's equations (1.1) and the rate equations (1.4) is achieved via $\vec{J} = \frac{d\vec{P}}{dt}$ in spatial regions occupied by molecules.

The system is excited by a plane wave at normal incidence generated using a total field/scattered field approach.[31] The periodic boundaries are imposed transversely to the incident field propagation. The absorbing boundaries in the form of convolutional perfectly matched layers (CPM) [35] are set on both sides of the array at a distance of 1 μm. The convergence of all spectra shown in figures 4 and 5 is achieved at the spatial resolution of 1.5 nm. The simulation domain varies depending on the period. The typical size of the domain is 250×250×1440. We parallelize our codes using message-passing interface (MPI) and propagate the Maxwell-Bloch equations on 72 processors at Thunder and Topaz SGI ICE X clusters. The typical execution time to obtain transmission/reflection spectra is 15 hours.

**Acknowledgements**

The experimental part is supported by the GIF grant proposal. The numerical modeling performed by M. S. is supported by the Air Force Office of Scientific Research under grant No. FA9550-15-1-0189 and the Binational Science Foundation under grant No. 2014113.

# Molecular Plasmonics: strong coupling at the low molecular density limit

Lihi Efremushkin,[a] Maxim Sukharev[b] and Adi Salomon[a]*

[a.] Department of Chemistry, Institute of Nanotechnology and Advanced Materials (BINA), Bar-Ilan University, Ramat-Gan 5290002, Israel.

[b.] Arizona State University, Mesa, AZ 85212, USA.

*Corresponding author E-mail: Adi.Salomon@biu.ac.il

| Absorbance | Concentration (M) | Density |
|---|---|---|
| 0.178 | 2.61E-03 | 1.57E+21 |
| 0.148 | 2.17E-03 | 1.31E+21 |
| 0.096 | 1.41E-03 | 8.48E+20 |
| 0.063 | 9.24E-04 | 5.56E+20 |
| 0.05 | 7.33E-04 | 4.41E+20 |
| 0.173 | 2.54E-03 | 1.53E+21 |
| 0.03 | 4.40E-04 | 2.65E+20 |
| 0.016 | 2.35E-04 | 1.41E+20 |

**Table S1**: The relation between the molecular absorbance and its concentration and density in the polymer. Given oscillator strength of $6.82 \times 10^6$ $M^{-1}cm^{-1}$ and based on Beer–Lambert law ($A=\varepsilon cl$), we calculate the molecular concentration (or density). The layer thickness is 100 nm, based on crossed section FIB.

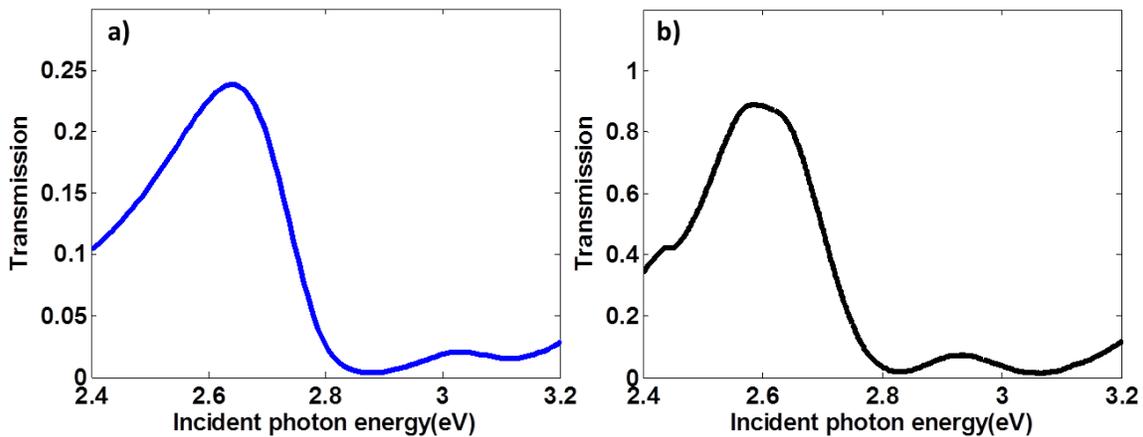

**Fig S1** (a) Transmission spectrum calculated for the periodic array of holes for the period of 330 nm. The hole diameter is 166 nm, the thickness of the silver is 150 nm and the PVA layer on the input side is 100 nm thick. The substrate is simulated as a semi-infinite, nondispersive dielectric with a refractive index of 1.5 corresponding to that of glass. The transmission is normalized to that in free space. (b) Transmission spectrum of periodic array of holes with the period of 360 nm covered with PVA. The hole diameter is 160 nm, the thickness of the silver is 250 nm. The transmission peak at 3.032 eV for the period of 330 nm corresponds to the experimental results for the 360 nm period

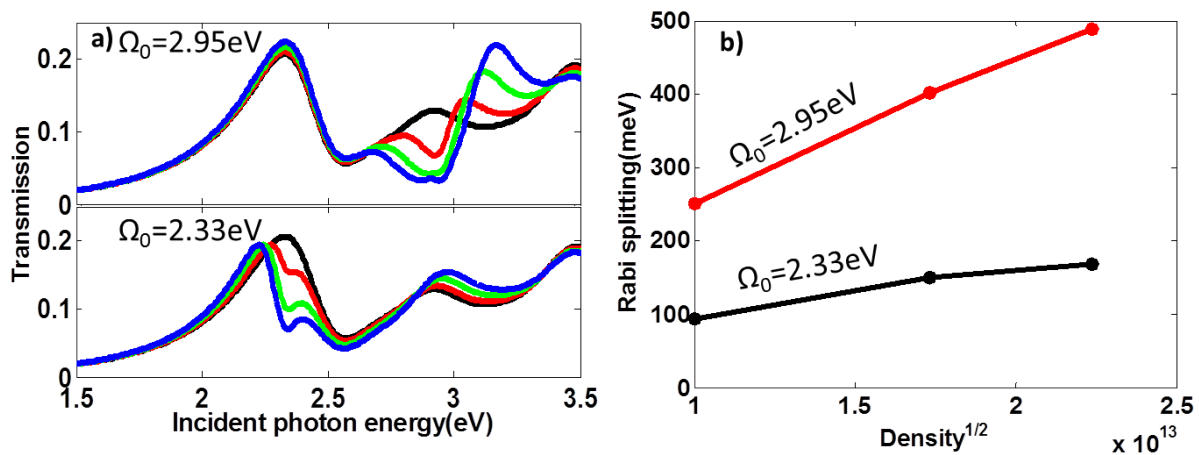

**Fig S2** (a) Transmission as a function of the incident photon energy calculated for the period of 330 nm. Transmission with no molecules in PVA is shown as a black line. Simulations with molecules doped in PVA for molecular concentrations are $10^{26}$ m$^{-3}$ (red line), $3\times10^{26}$ m$^{-3}$ (green line), $5\times10^{26}$ m$^{-3}$ (blue line). Other parameters used in the simulations are the molecular transition energy (2.33 eV), the transition dipole moment (10 Debye), the radiationless lifetime of the excited state ($4.14\times10^{-3}$ eV), the pure dephasing rate ($1.25\times10^{-2}$ eV), the PVA thickness (10 nm), the silver thickness (150 nm) and the hole diameter (166 nm). (b) Rabi splitting values for the simulated hybrid system presented in (a) $\Omega_0=2.95$ eV (red line) and 2.33 eV (black line).